\documentclass[aps,preprintnumbers,
				,nofootinbib]{revtex4}
\newcommand{\PRE}[1]{{#1}} 

\usepackage{graphicx}  
\usepackage{subfigure}
\usepackage{dcolumn}   
\usepackage{bm}        
\usepackage{amssymb}   
\usepackage{amsmath}   
\usepackage{ulem}
\usepackage{dcolumn}
\usepackage{epsfig}
\usepackage{array}
\usepackage{url}
\usepackage{setspace}
\usepackage[subfigure]{tocloft}
\newcommand{\gev}  {{\rm GeV}}

\newcommand{\tev}  {{\rm TeV}}

\newcommand{\invfb}{{\rm fb}^{-1}}

\newcommand{\etal} {{\it et~al.}}

\newcommand{\ptvis}  {p_{\rm T}^{\rm vis}}
\newcommand{\et}  {E_{\rm T}}

\newcommand{\met} {{E\!\!\!\!/_{\rm T}}}

\newcommand{\tanb} {\tan\beta}

\newcommand{\mzero}{m_{0}}
\newcommand{\mhalf}{m_{1/2}}

\newcommand{ \squark}   {\tilde{q}}

\newcommand{ \schionezero }{\tilde{\chi}_{1}^{0}}

\newcommand{ \schifourzero }{\tilde{\chi}_{4}^{0}}
\newcommand{ \schionepm }{\tilde{\chi}_{1}^{\pm}}
\newcommand{ \schionemp }{\tilde{\chi}_{1}^{\mp}}

\newcommand{ \alpgen } {{\tt ALPGEN}}
\newcommand{ \pythia } {{\tt PYTHIA}}

\newcommand{ \isajet }    {{\tt ISAJET}}
\newcommand{ \pgs }    {{\tt PGS4}}

\begin{document}

\preprint{MIFPA-11-12}

\title{
\PRE{\vspace*{1.3in}}
Bi-Event Subtraction Technique at Hadron Colliders
}

\author{Bhaskar Dutta$^1$}
\author{Teruki Kamon$^{1,2}$}
\author{Nikolay Kolev$^3$}
\author{Abram Krislock$^1$}

\affiliation{$^1$\ Department of Physics \& Astronomy,
Mitchell Institute for Fundamental Physics,
Texas A\&M University, College Station, TX 77843, USA
}
\affiliation{$^2$\ Department of Physics,
Kyungpook National University, Daegu 702-701, South Korea
}
\affiliation{$^3$\ Department of Physics, 
University of Regina, Regina, SK S4S 0A2, Canada
}

\begin{abstract}
\PRE{\vspace*{.3in}} We propose the Bi-Event Subtraction Technique (BEST) as a method of modeling and subtracting large portions of the combinatoric background during reconstruction of particle decay chains at hadron colliders. The combinatoric background arises when it is impossible to know experimentally which observed particles come from the decay chain of interest. The background shape can be modeled by combining observed particles from different collision events and be subtracted away, greatly reducing the overall background. This idea has been demonstrated in various experiments in the past. We generalize it by showing how to apply BEST multiple times in a row to fully reconstruct a cascade decay. We show the power of BEST with two simulated examples of its application towards reconstruction of the top quark and a supersymmetric decay chain at the Large Hadron Collider.
\end{abstract}

\maketitle

\bigskip
\newpage

The Large Hadron Collider (LHC) is up and running since 2009. Many models of particle physics beyond the Standard Model (SM) predict new particles which can be tested at the LHC. Heavy colored objects are expected to be produced at the LHC, followed by a chain of subsequent decays, according to such new models. Thus, we must fully or partially reconstruct these cascade decays from the particles which can be detected. However, reconstructions of these decays become experimentally difficult because it is impossible to know which particles come from the cascade decay we wish to reconstruct. The inevitable inclusion of particles which do not come from the cascade decay of interest is referred to as combinatoric background.

This combinatoric background can be removed easily in some cases by powerful subtraction techniques. For instance, the $Z$ boson can decay into oppositely charged, same flavored leptons: $Z\rightarrow e^+ e^- / \mu^+ \mu^-$. Leptons are easy to detect in the collider setting, and their charges can easily be measured. To reconstruct the $Z$ boson from these leptons, it is easy to collect a sample of Opposite-Sign Same-Flavor (OSSF) lepton pairs and construct the dilepton invariant mass for each pair. To model the combinatoric background, a sample of  Opposite-Sign Opposite-Flavor (OSOF) lepton pairs is selected as well. These OSOF lepton pairs cannot possibly both come from a single $Z$ boson, and so they model the combinatoric background well. Performing the OSSF$-$OSOF subtraction of the invariant mass distributions (possibly using some normalization factor $c$), $h^{\rm OSSF-OSOF}(m_{\ell \ell}) = h^{\rm OSSF}(m_{\ell \ell}) - c h^{\rm OSOF}(m_{\ell \ell})$, yields a distribution which shows a clear peak of the $Z$ boson mass.

However, such subtraction techniques are not available for jets, whose charges and flavors cannot so easily be determined. Thus, we introduce the Bi-Event Subtraction Technique (BEST) in which the combinatoric background of jets is modeled by combining jet information from a different event (or bi-event). This technique of modeling the combinatoric background by combining information from different events has been used before~\cite{bestOriginals}. However, here we generalize it, by applying it to jets. Moreover, we have shown that it can be used multiple times for the same decay chain reconstruction.

The basic idea of BEST can be demonstrated for the reconstruction of the $W$ boson decaying into two jets. For this case, a signal may be seen if a sample of jet pairs is collected for each event to construct the dijet invariant mass distribution, $h^{\rm same}(m_{jj})$. Here, the \textquotedblleft same\textquotedblright\ suggests that the jet pairs come from the same event. Some of the jet pairs in the same event distribution may come from a single $W$ boson decay in the events, while other jet pairs will be combinatoric background. By collecting another sample of jet pairs where each jet comes from a {\it different} event, the bi-event distribution, $h^{\rm bi}(m_{jj})$, can be formed. This bi-event distribution will have no jet pairs which come from a single $W$ boson. Thus, this bi-event distribution models a large amount of the combinatoric background well. The $h^{\rm bi}(m_{jj})$ distribution can be normalized to the $h^{\rm same}(m_{jj})$ distribution in the region of pure background (well away from the $W$ boson mass peak). For instance, the normalization factor can be calculated as
\begin{equation} 
	C_{jj}^{\rm BEST} 
		= \frac{\displaystyle\int_{150~{\rm GeV}}^{500~{\rm GeV}} h^{\rm same}(m_{jj}) {\rm d}m_{jj}}
			{\displaystyle\int_{150~{\rm GeV}}^{500~{\rm GeV}} h^{\rm bi}(m_{jj}) {\rm d}m_{jj}}.
	\label{eqBESTjjNormalization}
\end{equation} 
This normalization factor can be used when the shapes of these distributions are very close in this region. If the shapes of these distributions are not close, it could be due to some new physics. For instance, and additional resonance in the $h^{\rm same}(m_{jj})$ distribution could cause a mismatch in the shapes. However, it would be easy enough to recalculate the normalization taking an overall range which excludes the additional resonance. It should be noted that one needs a detailed systematic study of the shape from different physics processes. This is beyond the scope of this paper.

Finally, the BEST is performed: 
\begin{equation} 
	h^{\rm BEST}(m_{jj}) = h^{\rm same}(m_{jj}) - C_{jj}^{\rm BEST} h^{\rm bi}(m_{jj}).
	\label{eqBESTjj}
\end{equation}
The resulting dijet distribution shows a $W$ boson mass peak with most of the combinatoric background removed.

If we wish to reconstruct decay chains involving these $W$ bosons, we can take BEST even further. For instance, we can completely reconstruct the top quark from the decay chain $t\rightarrow bW\rightarrow bjj$. We can apply BEST again while combining the $b$ jets with the reconstructed $W$ bosons in order to reconstruct the top quark. However, this requires a more general application of BEST than has been used before.

For this example, we will refer to the same-event histograms by denoting the jets in the subscript as $j$ and $b$ for jets and $b$-jets respectively. For the bi-event histograms, we denote the jets in the subscript as $j'$ and $b'$. Thus we now denote our histograms and normalization factor from Eqs.~\eqref{eqBESTjjNormalization} and~\eqref{eqBESTjj} as:
\begin{subequations}
\begin{align}
	h^{\rm same}(m_{jj}) &\equiv h_{jj}(M_{jj}), \label{eqBESTnotation:a} \\
	h^{\rm bi}(m_{jj}) &\equiv h_{jj'}(M_{jj}), \label{eqBESTnotation:b} \\
	C^{\rm BEST}_{jj} &\equiv C^{\rm BEST\#1}_{jj}, \label{eqBESTnotation:c} \\
	h^{\rm BEST}(m_{jj}) &\equiv h^{\rm BEST\#1}_{jj}(m_{jj}) \label{eqBESTnotation:d}
\end{align}
\end{subequations}

To combine the reconstructed $W$ bosons with the $b$-jets to reconstruct the top quarks, we will need the following four additional histograms in order to perform two applications of BEST: $h_{bjj}(m_{bjj})$, $h_{bjj'}(m_{bjj})$, $h_{b'jj}(m_{bjj})$, and $h_{b'jj'}(m_{bjj})$. We perform the first BEST using the normalization factor calculated above in Eq.~\eqref{eqBESTjjNormalization}:
\begin{subequations}
\begin{align}
	h^{\rm BEST\#1}_{bjj}(m_{bjj}) = h_{bjj}(m_{bjj}) - C^{\rm BEST\#1}_{jj} h_{bjj'}(m_{bjj}) 
		\label{eqBESTnumberOne:a} \\
	h^{\rm BEST\#1}_{b'jj}(m_{bjj}) = h_{b'jj}(m_{bjj}) - C^{\rm BEST\#1}_{jj} h_{b'jj'}(m_{bjj}) 
		\label{eqBESTnumberOne:b}
\end{align}
\end{subequations}
Next we calculate another normalization factor for the second BEST which involves the combinatoric background of the $b$-jets. Once again, the range of this normalization factor is aimed at the region of pure background away from the top quark mass peak. Thus, it is calculated as:
\begin{equation} 
	C^{\rm BEST\#2}_{bjj}
		= \frac{\displaystyle\int_{200~{\rm GeV}}^{500~{\rm GeV}} 
				h^{\rm BEST\#1}_{bjj}(m_{bjj}) {\rm d}m_{bjj}}
			{\displaystyle\int_{200~{\rm GeV}}^{500~{\rm GeV}} 
				h^{\rm BEST\#1}_{b'jj}(m_{bjj}) {\rm d}m_{bjj}}.
	\label{eqBESTbjjNormalization}
\end{equation} 

With this normalization factor, we can finally perform the second BEST:
\begin{equation} 
	h^{\rm BEST\#2}_{bjj}(m_{bjj}) = h^{\rm BEST\#1}_{bjj}(m_{bjj}) 
			- C^{\rm BEST\#2}_{bjj} h^{\rm BEST\#1}_{b'jj}(m_{bjj}).
	\label{eqSecondBESTjj}
\end{equation}
Here, the resulting histogram will show a clean top quark mass peak with most of the combinatoric background removed. To clean up the resulting distribution even more, other subtraction techniques can also be employed, such as a sideband subtraction for the $W$ boson reconstruction. Each additional subtraction will double the number of initial histograms which are needed for all of the subtractions.

We demonstrate this powerful technique by using it to extract $W\rightarrow jj$ for (i) $t\bar{t}$ events at $\sqrt{s} = 7~\tev$ and (ii) SUSY events at $\sqrt{s} = 14~\tev$ within LHC simulations.

For the $t\bar{t}$ events, we generate hard scattering LHC collision events using \alpgen~\cite{alpgen}, perform the cascade decays with \pythia~\cite{pythia}, and perform a LHC detector simulation using \pgs~\cite{pgs}. The $W$+jets events are the main source of background for finding the top quark, so we generate these events in the same way. This background is mixed in randomly, according to production cross-sections, with our $t\bar{t}$ events. After \pgs\ is finished with these events, we select events for analysis with the following cuts~\cite{topMassCDF}: (i) Number of leptons, $N_\ell = 1$, where $p_T^{(\ell)} \ge 20~\gev$ and $p_{T,iso}^{(\ell)} \le 0.1\times p_T^{(\ell)}$; (ii) Missing transverse energy, $\met \ge 20~\gev$; (iii) Number of jets, $N_j \ge 3$, where $p_T^{(j)} \ge 30~\gev$ and at least one jet has been tightly $b$-tagged~\cite{pgs}; (iv) Number of taus, $N_\tau = 0$ for taus with $p_T^{(\tau)} \ge 20~\gev$~\cite{pgs}.

\begin{figure} 
	\includegraphics[width = 7.5cm]{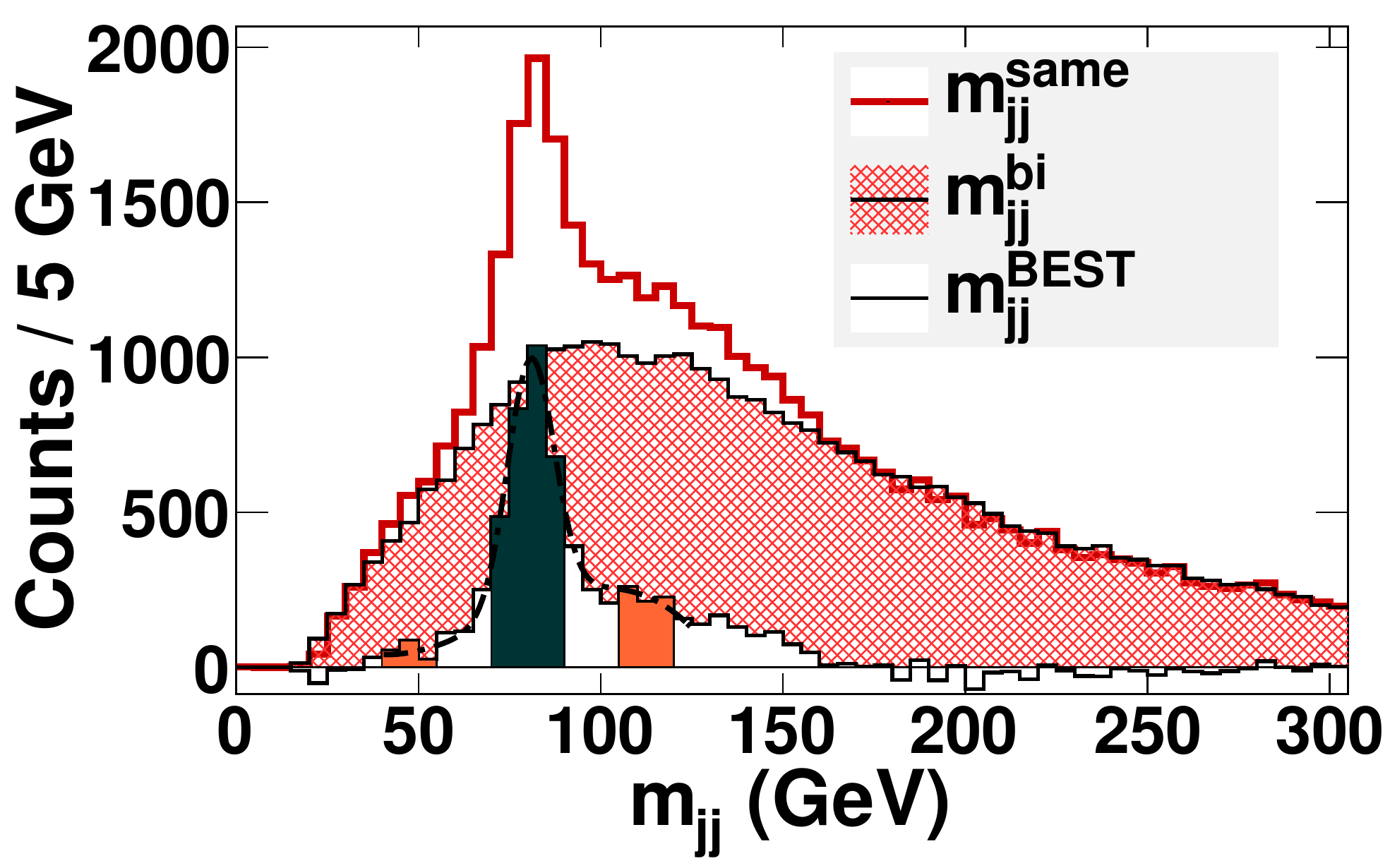}
	\caption{The dijet invariant mass distribution, $m_{jj}$. This plot shows the same-event ($m_{jj}^{\rm same}$), bi-event ($m_{jj}^{\rm bi}$), and BEST ($m_{jj}^{\rm BEST}$) distributions as described in the text. The BEST distribution is fitted with a gaussian plus cubic function, to find the $W$ boson mass peak and surrounding background. The BEST distribution is also split up into regions for a sideband subtraction used for reconstructing an invariant mass between a $W$ boson and a $b$ tagged jet. The $W$ region is dark cyan filled, while the sidebands are orange filled. For an integrated luminosity of $2~\invfb$, we find the $W$ boson mass, $m_W = 81.11 \pm 0.32~\gev$.}
	\label{figTTbarBESTfindsW}
\end{figure}

With our events selected in this way, we pair up jets (which are not $b$-tagged) to fill the same-event and bi-event $h(m_{jj})$ distributions as described above. Each jet pair must have $\Delta R \ge 0.4$. To fill the bi-event distribution, we refer to jets from the previous event which has passed the same cuts as listed above. Once the distributions are filled with all events, we normalize the shape of the $h^{\rm bi}(m_{jj})$ distribution as described by Eq.~(\ref{eqBESTjjNormalization}). Then we perform our BEST. The result of this subtraction can be seen in Fig.~\ref{figTTbarBESTfindsW}, which shows a drastic reduction in the background obscuring the $W$ boson reconstruction. Note that the bi-event distribution models the combinatoric background of any jet pairs which are not correlated by decay chains or event kinematics. Thus, BEST in this case removes (i) the combinatoric background from events with $W$ bosons (coming from $t$ decays) and (ii) uncorrelated jet pairs coming from our $W$+jets background sample.

\begin{figure} 
	\includegraphics[width = 7.5cm]{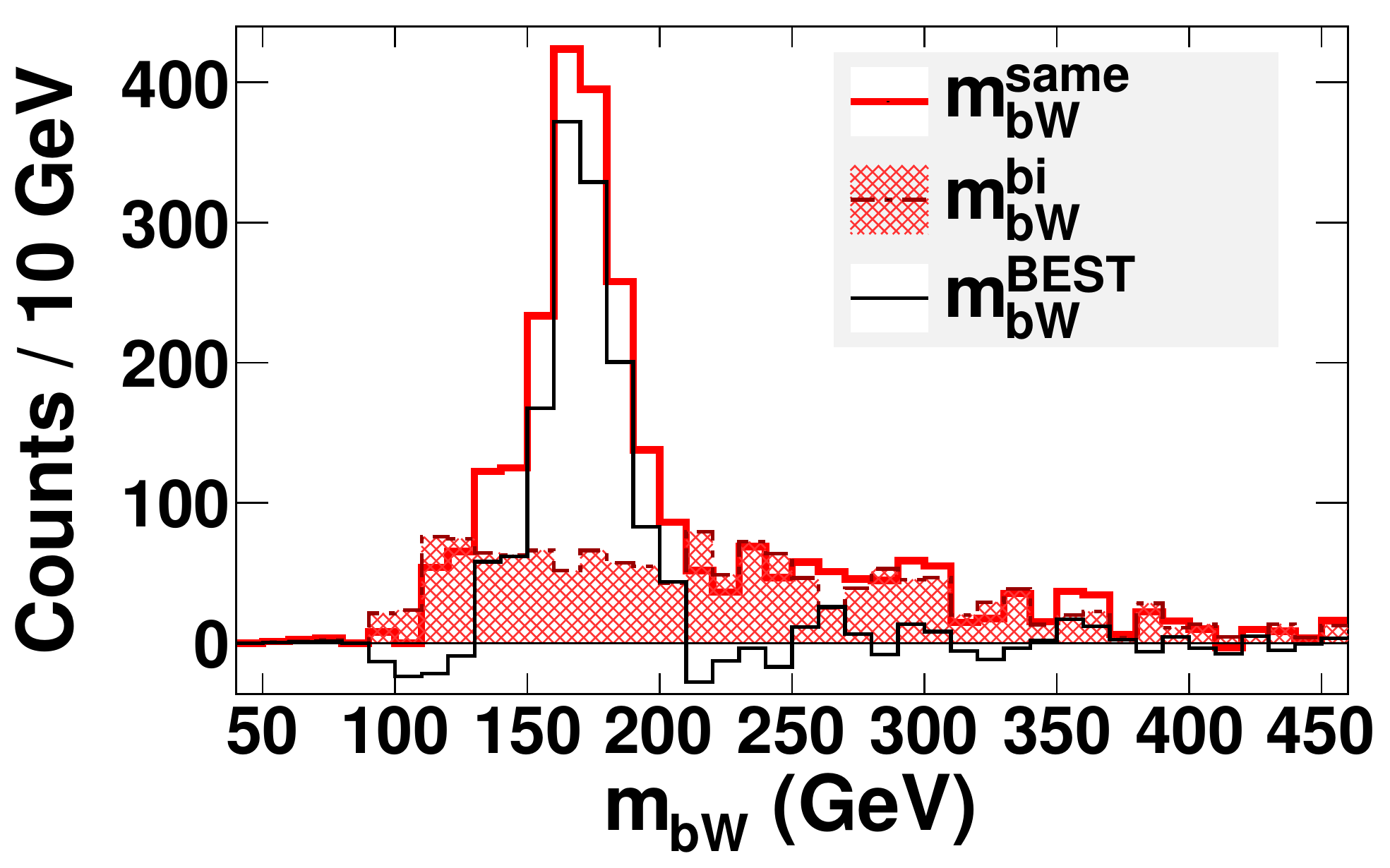}
	\caption{The $W$ plus $b$ invariant mass distribution, $m_{bW}$. This plot shows the same-event, bi-event, and BEST distributions as described in the text. For an integrated luminosity of $2~\invfb$, we find the top quark mass, $m_t = 170.5 \pm 1.5~\gev$. The top quark mass is set within \alpgen\ as $m_t = 174.3~\gev$.}
	\label{figTTbarBESTfindsTop}
\end{figure} 

\vspace{0.2cm}
Once we have found the $W$ boson with this first application of BEST, we can combine the $W$ boson with a $b$-jet to find the top quark. To remove additional background from the $W$ signal, we perform a sideband subtraction. To do this, we split up the dijet signal into a $W$ boson mass region, where $70~\gev \le m_{jj} \le 90~\gev$, and two sideband regions, $40~\gev \le m_{jj} \le 55~\gev$ and $105~\gev \le m_{jj} \le 120~\gev$. We form the dijet ($W$) plus $b$ invariant mass, keeping track of whether the dijet system was in the $W$ window or sideband windows. In this way we make the $W$ band ($h_{W{\rm band,\ BEST}}(m_{bW})$) and sideband ($h^{\rm SB,\ BEST}(m_{bW})$) distributions. The sideband distribution models the remaining background of $W$'s very well. By fitting the $h^{\rm BEST}(m_{jj})$ by a gaussian function, $f(m_{jj}^{\rm BEST})$, plus a background function, $g^{\rm BG}(m_{jj}^{\rm BEST})$, we can find the shape of the background distribution which remains. Then we calculate a normalization factor:
\begin{equation}
	C_{jj}^{\rm SB} 
		= \frac{\displaystyle\int_{W {\rm band}}
				g^{\rm BG}(m_{jj}^{\rm BEST}) {\rm d}m_{jj}^{\rm BEST}}
			{\displaystyle\int^{\rm SBs} 
				g^{\rm BG}(m_{jj}^{\rm BEST}) {\rm d}m_{jj}^{\rm BEST}},
	\label{eqSidebandjjNormalization}
\end{equation}
Using this normalization factor, we perform the sideband subtraction, 
\begin{equation}
	\begin{tabular}{l}
	$h^{\rm SBsub,\ BEST}(m_{bW}) =$ \\
		\hspace{0.3cm}$h_{W{\rm band,\ BEST}}(m_{bW}) - C_{jj}^{\rm SB} h^{\rm SB,\ BEST}(m_{bW}).$
	\end{tabular}
	\label{eqSidebandjj}
\end{equation}
This subtraction removes even more of the $W$ combinatoric background. 

Lastly, to remove the combinatoric background of $b$-jets, we can perform our BEST again. We form the $h^{\rm SBsub,\ BEST}(m_{bW})$ distribution again, this time using $b$ jets which come from a different event as the $W$. Again, this models the combinatoric background very well, since the $W$ and $b$ from different events cannot possibly come from a single top quark. We can calculate a normalization factor as before analogous to Eq.~(\ref{eqBESTjjNormalization}) in the range $200~\gev \le m_{bW} \le 500~\gev$ (a little away from the top mass peak). Using this normalization factor, we can perform the final BEST, analogously to that shown in Eq.~(\ref{eqSecondBESTjj}). The resulting $m_{bW}$ distribution after this last application of BEST is shown in Fig.~\ref{figTTbarBESTfindsTop}, which shows a very clean looking top peak.

In the context of top reconstruction, other groups have come up with some techniques to eliminate the combinatoric background. In experimental top reconstruction~\cite{topMassCDF,topMassCMS,topMassATLAS}, combinatoric background is eliminated by assuming a very particular event topology. By selecting certian events, the combinatoric background is eliminated by essentially choosing the jet combinations which form the best $W$ and $t$ masses. For SM $t\bar{t}$ events, this reconstruction works quite well to measure the top mass. However, these methods cannot be employed to reconstruct $t$ quarks from beyond SM sources. On the other hand, some phenomenological studies of top production from beyond SM use top tagging~\cite{topTagger} to identify the top correctly, and thus reduce the combinatoric background. However, the top tagging relies on the production of a boosted top from the decay of a heavy new particle. Also, although this top tagger has a large efficiency, it seems the fake rate from SM backgrounds may be large. These experimental and phenomenological techniques may be more precise than BEST (although, a thorough study would be needed to compare them). However, the advantage of BEST is that it does not require any assumptions about the event topology or having boosted tops.

This example from the SM shows the power of BEST. Additionally, BEST is useful for searches and measurements of models beyond the SM. Thus, we also demonstrate the use of BEST for a supersymmetry (SUSY) model. The model we choose is the non-universal generalization of the minimal supergravity model~\cite{msugra} i.e., nuSUGRA. In this nuSUGRA model, the Higgs masses are not unified with the other scalar masses at the grand unified scale. This allows for a more general mass spectrum than that of the mSUGRA model. The indication of the preference for the nuSUGRA model at the LHC is that the neutralino masses will not have the mass ratios predicted by mSUGRA. This nuSUGRA model can also predict the correct amount of dark matter in the universe today. In particular, a large parameter space region of this model has an abundance of $W$ bosons being produced~\cite{nuSUGRAatLHC}. These $W$ bosons must be found and utilized to reconstruct the model. Thus, this is a useful model to examine with BEST.

\begin{figure} 
	\includegraphics[width = 7.5cm]{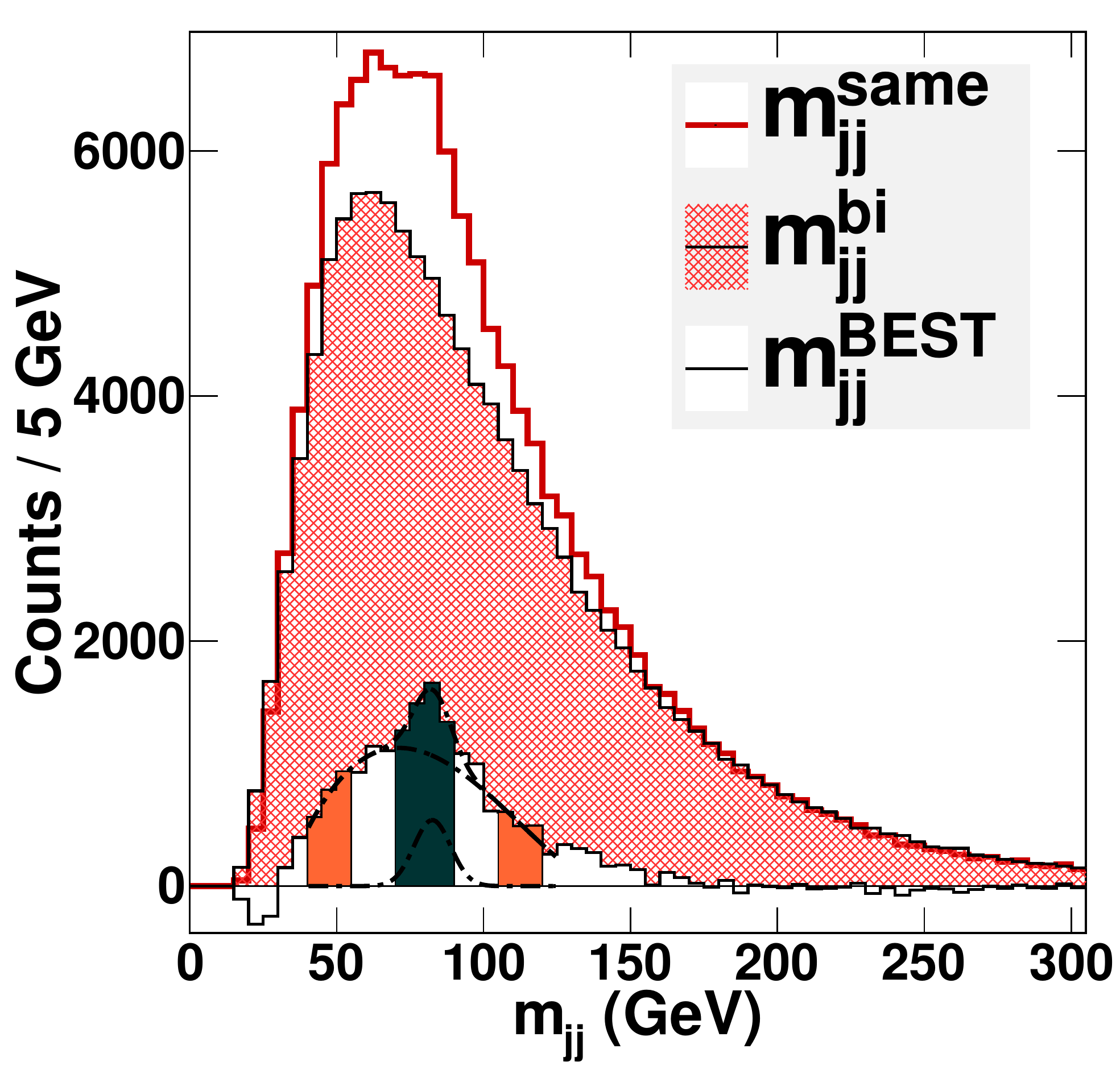}
	\caption{The dijet invariant mass distribution, $m_{jj}$ from our nuSUGRA events mixed with SM backgrounds. The BEST has already been performed. The BEST distribution is fitted and split up into regions for a sideband subtraction used for reconstructing an invariant mass between a $W$ boson and a leading jet. The $W$ region is dark cyan filled, while the sidebands are orange filled. Here we find the $W$ boson mass, $m_W = 82.4 \pm 1.0~\gev$. This plot is for an integrated luminosity of $100~\invfb$.}
	\label{fignuSUGRABESTfindsW}
\end{figure} 

We choose a benchmark point for the nuSUGRA model for this demonstration: $\mzero = 360~\gev$, $\mhalf = 500~\gev$, $\tanb = 40$, $A_0 = 0$, and $m_H = 732~\gev$, with the top mass set as $m_t = 172.6~\gev$. This point in parameter space predicts an abundance of $W$ bosons at the LHC due to neutralino or chargino decays. The decay chain we wish to partially reconstruct is: $\squark \rightarrow q + \schionepm \left(\schifourzero\right) \rightarrow q + W^{\pm} + \schionezero \left(\schionemp\right)$. Our BEST has been used to analyze this signal already, with the details shown in \cite{nuSUGRAatLHC}.

To simulate events for this demonstration, we once again use \pythia\ and \pgs. The SUSY mass spectrum is generated using \isajet~\cite{isajet}. We also use \alpgen\ to simulate some SM backgrounds. The primary SM backgrounds for the events we wish to analyze are $Z$+jets, $W$+jets, and $t\bar{t}$ events. We mix these SM backgrounds in randomly with our SUSY signal events.

\begin{figure} 
	\includegraphics[width = 7.5cm]{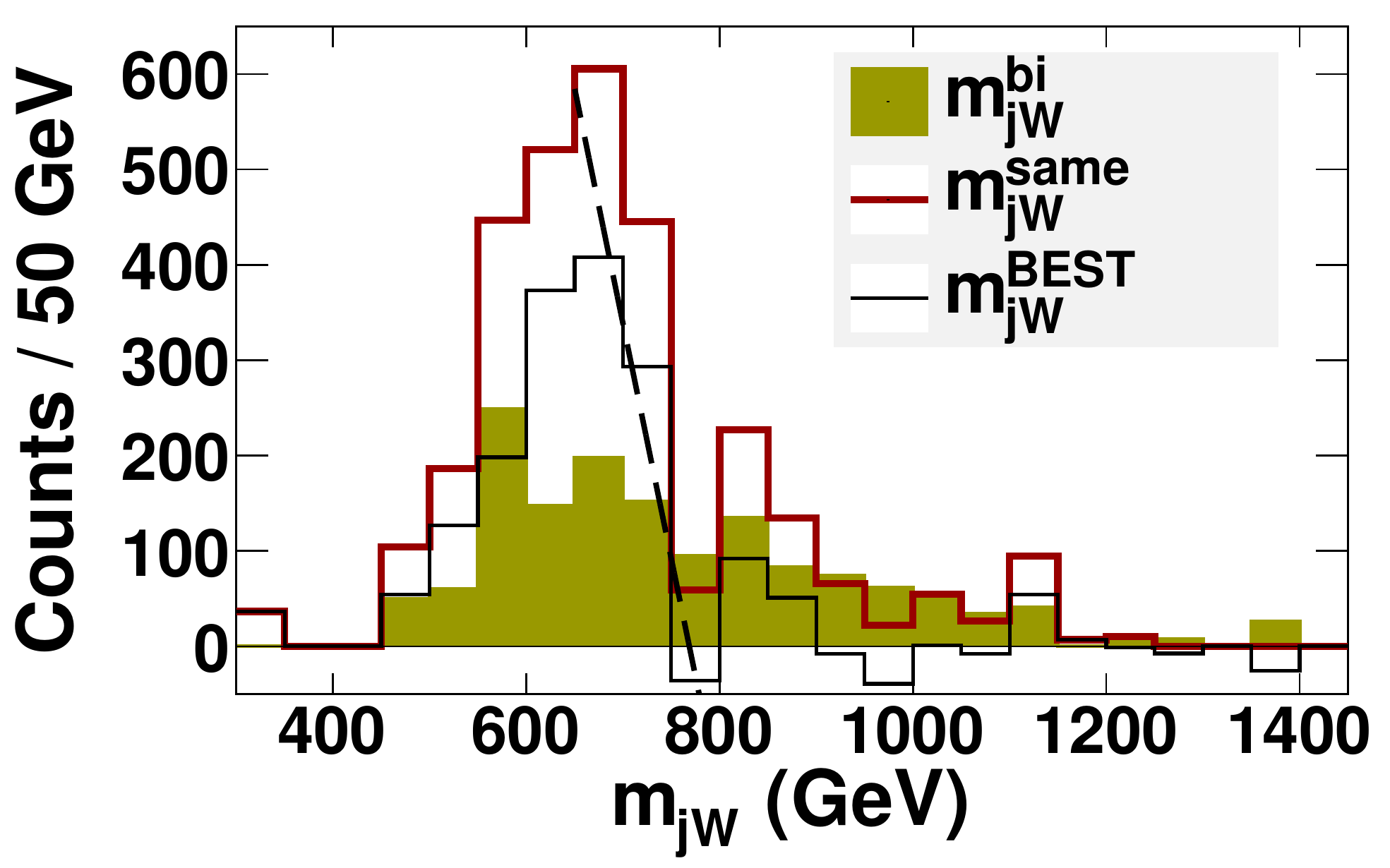}
	\caption{The $W$ plus jet invariant mass distribution, $m_{jW}$. This plot shows the same-event, bi-event, and BEST distributions as described in the text. BEST removes the background obscuring the endpoint. For an integrated luminosity of $100~\invfb$, we find the endpoint to be $769 \pm 18~\gev$. This is within $2\sigma$ of the theoretical endpoint, which is $738.8~\gev$ for the most probable decay chain of this type, $\squark \rightarrow q + \schifourzero \rightarrow q + W^{\pm} + \schionemp$.}
	\label{fignuSUGRABESTfindsEndpoint}
\end{figure} 

\vspace{0.2cm}
To help reduce the SM backgrounds, we use the following selection cuts, which are refined from the cuts in \cite{nuSUGRAatLHC}: (i) Missing transverse energy, $\met \ge 180~\gev$; (ii) Number of jets, $N_j \ge 4$, where $p_T^{(j)} \ge 30~\gev$; (iii) Minimum $\Delta\phi$ between leading three jets and missing transverse energy, $\Delta\phi^{\rm min} \ge 0.5$; (iv) Leading jet transverse momenta, $p_T^{({\rm 1st}\ j)} \ge 300~\gev$ and $p_T^{({\rm 2nd}\ j)} \ge 200~\gev$; (v) $\Delta R$ between leading jets, $\Delta R({\rm 1st\ }j, {\rm 2nd\ }j) \le 3.2$; (vi) Scalar sum, $p_T^{({\rm 1st}\ j)} + p_T^{({\rm 2nd}\ j)} + 3\cdot\met \ge 1600~\gev$.

With these event selection cuts, we begin to pair up the sub-leading jets as we did for the $t\bar{t}$ analysis, perform the BEST to find the $W$ bosons, then combine the $W$'s with the leading jets to reconstruct the desired decay chain. While pairing up the jets, we use the additional cut $0.4 \le \Delta R(jj) \le 1.5$. We once again perform a sideband subtraction to help clean up any excess background involved with finding the $W$ bosons. When combining the $W$ candidates (jet pairs) with leading jets, we keep only those combinations where $\Delta R(W, j) \ge 1.0$. We use BEST again on the leading jet as well, to remove combinatoric background from the leading jets which are not from our desired decay chain. The result of this analysis can be seen in Figs.~\ref{fignuSUGRABESTfindsW} and~\ref{fignuSUGRABESTfindsEndpoint}. Note in Fig.~\ref{fignuSUGRABESTfindsW} that the $W$ boson mass peak can barely be seen in the same-event histogram, but is clearly visible after the application of  BEST.

In conclusion, BEST is a powerful subtraction technique which can find and reconstruct particles normally hidden by the combinatoric background, as shown in Fig.~\ref{fignuSUGRABESTfindsW}.
It is useful for the further understanding of the SM as well as models beyond the SM. It can be utilized without information about the charge or flavor of the particles involved. BEST can, therefore, improve any current and future collider study and help us detect new particles, measure their masses and determine model parameters accurately.

\section*{Acknowledgements}
This work is supported in part by the DOE grant DE-FG02-95ER40917 and by the World Class University (WCU) project through the National Research Foundation (NRF) of Korea funded by the Ministry of Education, Science \& Technology (grant No. R32-2008-000-20001-0). We would like to thank D. Feldman, I. Hinchilffe, and S. Su for useful discussions, and Y. Oh for providing computer support.

\end{document}